\documentclass[citeautoscript,floatfix,aps,prb,twocolumn,superscriptaddress]{revtex4-2}
\usepackage{graphicx}
\usepackage{dcolumn}
\usepackage{amsmath}
\usepackage{amssymb}
\usepackage{bm}
\usepackage{lipsum} 
\usepackage{graphicx} 
\usepackage[font=small,labelfont=bf,justification=raggedright,format=plain]{caption} 
\usepackage{tikz}
\usepackage{array}
\usepackage{microtype}
\usepackage{relsize}
\usepackage{MnSymbol}
\usepackage{xcolor}
\usepackage{wasysym}
\usepackage{caption}
\usepackage{subcaption}
\usepackage{comment}
\usepackage{url}
\usepackage{hyperref}
\usepackage{tikz-feynman}
\hypersetup{
	colorlinks=true,
	final=true,
	linkcolor=blue,
	filecolor=magenta,
	urlcolor=blue,
	citecolor= blue,
	bookmarks=true,
}

\begin{document}
	
	\title{Inhomogeneous phase stiffness in two-dimensional $s$-wave disordered
		superconductors}

\author{Sudipta Biswas}
\email{sbiswas4396@iitkgp.ac.in}
\affiliation{Department of Physics, Indian Institute of Technology Kharagpur, Kharagpur 721302, West Bengal, India}

\author{A. Taraphder}
\email{arghya@phy.iitkgp.ac.in}
\affiliation{Department of Physics, Indian Institute of Technology Kharagpur, Kharagpur 721302, West Bengal, India}

\author{Sudhansu S. Mandal}
\email{sudhansu@phy.iitkgp.ac.in}
\affiliation{Department of Physics, Indian Institute of Technology Kharagpur, Kharagpur 721302, West Bengal, India}

\begin{abstract}
	We investigate the effect of white-noise disorder on the local phase stiffness and thermodynamic properties of a two-dimensional $s$-wave superconductor. Starting from a local attractive model and using path-integral formalism, we derive an effective action by decoupling the superconducting order parameter into amplitude and phase components in a gauge-invariant manner. Perturbative techniques are applied to the phase fluctuation sector to derive an effective phase-only XY model for disordered superconducting systems. Solving the saddle-point Green's function using Bogoliubov-de Gennes theory, we calculate the distributions of nearest-neighbor couplings for various disorder strengths. A single-peak distribution is observed for low disorder strength, which becomes bimodal with one peak at negative couplings as the disorder strength increases. The local phase stiffness remains randomly distributed throughout the lattice and shows no correlation with pairing amplitudes. The temperature dependence of the superfluid stiffness ($J_s$) is studied using Monte Carlo simulations. At strong disorder and low temperatures, $J_s$ increases with increasing temperature, exhibiting anomalous behavior that may indicate the onset of a glassy transition. Additionally, calculations of the Edwards-Anderson order parameter in this disorder regime suggest the emergence of a $phase$-$glass$ state at very low temperatures.
\end{abstract}

\maketitle

\section{\label{sec:I}INTRODUCTION}

The interplay between superconductivity and localization in two dimensions is a longstanding and complex problem. Early work by Anderson \cite{Anderson_1958, Anderson_1959} addressed the low-disorder regime, where the mean free path is significantly longer than the inverse Fermi momentum. The critical temperature remains unchanged in this regime, consistent with the eponymous theorem. However, in the moderate to high disorder regime, where it no longer applies, the localization of the superconducting (SC) state leads to a superconductor-to-insulator transition (SIT). This transition has been extensively studied over the past few decades through both theoretical \cite{Fisher_1990, Sauls_2022, Beloborodov_2007,  Ghoshal_2001, Avishai_2007} and experimental \cite{Goldman_1989, Pratap_2012, Pratap_2013} investigations. Nevertheless, the microscopic mechanisms driving the SC state to an insulating one and its nature remain unresolved.

The superfluid stiffness, also referred to as the global phase stiffness \( J_s \), quantifies the energy cost associated with imposing a small uniform phase gradient across the system, whereas the local phase stiffness or local coupling \( J_{ij} \) quantifies the energy cost for a relative phase twist between neighboring sites \( i \) and \( j \). Physically, \(J_s\) characterizes the superconducting system's response to an externally applied transverse gauge field. For conventional clean superconductors, $J_s$ is a much larger energy scale than SC pairing amplitude $\Delta$, making phase fluctuation irrelevant \cite{Pratap_2022}. In disordered superconductors, however, $J_s$ could become smaller than $\Delta$, resulting in phase-driven phenomena. It has been observed experimentally \cite{Mintu_2011,Sherman_2012,Noat_2013} that thin films of conventional superconductors at the brink of SIT show a finite $\Delta$ above $T_c$, suggesting a phase transition driven by phase coherence\cite{Ioffe_2010,Feigelman_2010,Seibold_2012,Lemarie_2013}. This is attributed to the fact that near SIT, the configuration-averaged superfluid density, which is proportional to superfluid stiffness at zero frequency and momentum \cite{Scalapino_1993,AT_1996}, becomes small and eventually vanishes at the transition. However, the configuration-dependent local superfluid density fluctuates extensively from site to site both in sign and magnitude \cite{Kivelson_1991,Kivelson_1992}. Therefore, at strong disorder, fluctuation leads to emergent inhomogeneity in the SC ground state, suggesting a granular SC landscape. This calls for a mapping of the SC problem onto an effective phase-only model (an XY model), where the disorder is encoded in the local couplings $J_{ij}$ \cite{Maccari_2017,Maccari_2019,Nandini_2014}.

In 2D, Berezinskii-Kosterlitz-Thouless (BKT) theory \cite{Berezinsky_1972,Kosterlitz_1973,Kosterlitz_1974} describes the superfluid-to-normal transition; shows universal jump of the superfluid density at the transition \cite{Nelson_1977}. A previous study of the temperature dependence of superfluid stiffness in the disordered XY model provides an effective theory for describing low-temperature stiffness using a perturbative expansion in the disorder potential \cite{Maccari_2019}. This analytical approach offers a good description of the Monte Carlo results for both the Gaussian and diluted models of disorder at low temperature. However, in the presence of spatially correlated coupling \cite{Maccari_2017}, the universal sharp jump in stiffness is substantially smeared out. This effect is attributed to a different mechanism of vortex-antivortex pair generation, driven by the formation of large clusters of weak superconducting regions, as observed experimentally \cite{Mintu_2011, Kamlapure_2010}. All these studies begin with a proper phase-only model, e.g. classical XY model, considering two prototypes of distributions of local couplings: \textbf{i)} Gaussian model, characterized by a distribution of local phase stiffnesses centered around a mean value, describes a weak disorder regime. \textbf{ii)} Diluted model, where the local coupling can be either zero or a specific finite mean value, depending on the dilution parameter, corresponds to a relatively strong disorder regime. These works motivate us to investigate if a microscopic formalism, which yields these kinds of distributions, is possible.

Starting from a locally attractive Hamiltonian in the presence of white-noise disorder, we use the path integral approach to write down the partition function in terms of an effective action, employing the Hubbard-Stratonovich transformation. Breaking the effective action into two parts in a gauge invariant manner, one is the mean-field part, also known as the amplitude part and the other one, the phase fluctuation part, which is expanded up to quadratic order (Gaussian or harmonic approximation)\cite{TVR_1989,SSM_2020}. Mean-field part is solved numerically by the mean-field Bogoliubov-de Gennes (BdG) formalism \cite{de_gennes_1966,SSM_2013,Mohanta_2013,Ravi_2024} to get the eigenvalues and eigenvectors. Considering only the spatial dependence of the superconducting phase, we express the fluctuation action in a form analogous to the classical XY model Hamiltonian: $ H_{XY} = - \sum_{\langle ij \rangle} J_{ij} \cos(\theta_{i} - \theta_{j})$. Here, the local Josephson coupling \( J_{ij} \) is derived by expanding the phase action up to Gaussian order and is computed numerically using mean-field BdG solutions. We see that at low disorder strength, the distribution exhibits a single peak with a shape approximately resembling a Gaussian. As the disorder increases further, a secondary peak starts to appear at a finite negative value and at substantially high disorder – the distribution becomes bimodal. This exhibits a peak on the negative side, representing an unconventional feature that deviates from the behavior assumed in previously proposed diluted models of local couplings (two-delta distribution) \cite{Maccari_2019}. The occurrence of negative local coupling is a significant deviation from conventional behavior. A finite admixture of negative and positive couplings drives the system into a frustrated phase. This leads us to investigate the possibility of a $glass$-like state at high disorder.

We further investigate the temperature dependence of stiffness for different disorder strengths by Monte Carlo (MC) simulation in a classical XY model. At low disorder, the universal sharp jump of superfluid stiffness was observed at the critical temperature, which is the hallmark of the BKT transition in two dimensions. At moderate to high disorder strength, however, we observe smearing of this jump, in agreement with the systematically broadened jumps observed experimentally in thin films of conventional superconductors \cite{Mintu_2011,Pratap_2012,Pratap_2013}. Apart from the smearing of the BKT jump, it was observed that disorder could affect the low-temperature behavior of the superfluid stiffness in a nontrivial way, depending on the probability distribution of local stiffness. At high disorder strength, the low-temperature behavior of the superfluid stiffness for a bimodal-like \( J_{ij} \) distribution exhibits nontrivial features indicative of a phase-glass-like state emerging beyond a critical disorder threshold. To support this interpretation, we compute the Edwards-Anderson (EA) order parameter~\cite{Edwards_1975,Parisi_1983}, which confirms the presence of glassy order. These results agree with previous theoretical studies on spin-glass systems~\cite{Halsey_1985,Dasgupta_1988}.

The paper is organized as follows: In Sec. \ref{sec:II}, we describe the model Hamiltonian for studying disordered superconductors in a lattice model and use the path-integral formalism to segregate the mean field (amplitude-dependent part) and phase-dependent part of effective action. We then present a method to compute the nearest-neighbor couplings from the phase-dependent part of the action. Section \ref{sec:III} describes the characteristics of the distributions of local phase stiffness for various disorder strengths, along with a lattice bond plot to examine correlations with other local properties. We then calculate the temperature dependence of the superfluid stiffness along with diamagnetic and paramagnetic responses using the MC simulation in a classical XY model and finally find critical temperatures for various disorder strengths. Later, we confirm the possibility of a glassy state by calculating EA order parameter with temperature. Finally, in Sec. \ref{sec:IV}, we conclude with the key feature of our calculations and results. Appendix~\ref{app:A} provides a detailed derivation of the second-order contributions to the local couplings from the phase action. In Appendix~\ref{app:B}, we show the connection between the saddle point of the effective action and BdG mean-field theory, as well as the form of Nambu Green's function in the Matsubara frequency domain in terms of BdG eigenvectors.

\section{\label{sec:II}Formalism and Theory}
	
	 We begin with a nearest neighbor tight-binding model Hamiltonian for electrons in a square lattice having onsite disorder potentials $V_i$ and negative-$U$ Hubbard interactions between electrons. The corresponding Hamiltonian is given by
	\begin{equation}\label{eqn1}
		\mathcal{H} = - t \sum_{<ij>,\sigma} (c^\dagger_{i\sigma}c_{j\sigma} + h.c.) - \mu \sum_{i,\sigma}n_{i\sigma} +  \sum_{i,\sigma}V_i n_{i\sigma} - U \sum_i n_{i\uparrow} n_{i\downarrow} \, .
	\end{equation}
	Here $c^\dagger_{i\sigma}\,(c_{i\sigma})$ is the electron creation (destruction) operator with spin-$\sigma\, (\uparrow,\downarrow)$ at $i^{\rm th}$ site, t is the hopping energy,  $    n_{i\sigma} = c^\dagger_{i\sigma}c_{i\sigma}$ is the number operator for spin-$\sigma$, and $\mu$ is the chemical potential for a system of fixed electron density. In the coherent-state path-integral approach, the partition function of the system can be expressed in terms of integrals of Grassmanian variables $c_{i\sigma}$ and $\bar{c}_{i\sigma} $ corresponding to fermionic  destruction and creation operators respectively as 
	\begin{equation}\label{eqn2}
		\mathcal{Z} = \int \mathcal{D}[\bar{c}_{\sigma}, c_{\sigma}] e^{-\mathcal{S}[\bar{c}_{\sigma}, c_{\sigma}]}
	\end{equation}
	with the Euclidean action for imaginary time $\tau$, 
	{\small
	\begin{eqnarray}\label{eqn3}
	&&	\mathcal{S} =  \int_{0}^{\beta} d\tau \left[
		\sum_{i,\sigma} \bar{c}_{i\sigma}(\tau) [\partial_{\tau}+ (V_i-\mu) ] c_{i\sigma}(\tau)\right. \nonumber \\ 
	&&	 \left. -t\sum_{\langle ij\rangle,\sigma} \bar{c}_{i\sigma}(\tau)  c_{j\sigma}(\tau)
		 - U \sum_i \bar{c}_{i\uparrow}(\tau) \bar{c}_{i\downarrow}(\tau) c_{i\downarrow}(\tau) c_{i\uparrow}(\tau)\right]
	\end{eqnarray}
	}
	where $\beta = 1/(k_BT)$, T being the temperature of the system. By introducing complex scalar variable $\varphi_i(\tau)$ and its conjugate $\varphi^\ast_i(\tau)$, and a real scalar variable $\xi_i(\tau)$, we perform Hubbard-Stratonovich transformation for finding terms with bilinear Grassmanian variables rather than the biquadratic term in $\mathcal{S}$, we find
	\begin{equation}\label{eqn4}
		\mathcal{Z} = \int \mathcal{D}[\bar{c}_{\sigma}, c_{\sigma}] \mathcal{D} [\varphi^\star,\varphi] \mathcal{D} [\xi] e^{-\mathcal{S^{\prime}}[\bar{c}_{\sigma}, c_{\sigma}, \varphi^\star,\varphi , \xi]}
	\end{equation}
	where,
	\begin{eqnarray}\label{eqn5}
		&& \mathcal{S^{\prime}} =  \int_{0}^{\beta} d\tau \biggl[\mathlarger{\sum_i}\biggl(\dfrac{|\varphi_i(\tau)|^2}{U}+\dfrac{\xi^2_i(\tau)}{2U}\biggr) \nonumber \\
		&& +\sum_{i,\sigma} \bar{c}_{i\sigma}(\tau) [\partial_{\tau}+ (V_i-\tilde{\mu}_i) ] c_{i\sigma}(\tau) -t \sum_{\langle ij\rangle,\sigma} \bar{c}_{i\sigma}(\tau)  c_{j\sigma}(\tau) \nonumber \\
		&& + \sum_i [\varphi_i(\tau) \bar{c}_{i\uparrow}(\tau) \bar{c}_{i\downarrow}(\tau) + \varphi^{\star}_i(\tau)c_{i\downarrow}(\tau) c_{i\uparrow}(\tau)  ] \biggr]
	\end{eqnarray} \\
	where local effective chemical potential becomes $\tilde{\mu}_i = \mu + \xi_i $. In analogy to BCS Hamiltonian, $\varphi_j(\tau)$ is identified as the superconducting complex order parameter consisting of amplitude and phase: $\varphi_j(\tau) = \Delta_j(\tau) e^{i\theta_j(\tau)}$. We next make an asymmetric gauge transformation, $ c_{i\uparrow}(\tau) \rightarrow c_{i\uparrow}(\tau)  e^{i\theta_i(\tau)})$, $ c_{i\downarrow}(\tau) \rightarrow c_{i\downarrow}(\tau)$ for gauged away the phase factor of the order parameter. This leads to the form of $\mathcal{S}'$ as
	{\small
	\begin{eqnarray}\label{eqn6}
		&& \mathcal{S^{\prime}} = \int_{0}^{\beta} d\tau \hspace{0.2em}\biggl[ 
		\mathlarger{\sum_i} \biggl(\dfrac{|\varphi_i(\tau)|^2}{U}+\dfrac{\xi^2_i(\tau)}{2U}\biggr) \nonumber \\
		  && +  \mathlarger{\sum_{i,j}} \bar{\Psi}_{i}(\tau) \biggl \{ \biggl( \sigma_0 \partial_{\tau} - \tilde{\mu_i}\sigma_3 + V_i\sigma_3 + \Delta_i \sigma_1 + i\dot{\theta}_i(\tau) \dfrac{\sigma_0+\sigma_3}{2} \biggr)\delta_{ij} \nonumber \\
		&& - t\sigma_3
		\mathlarger{\sum}_{\bm{\delta}=\pm \hat{x},\pm \hat{y}} \delta_{j,i+\bm{\delta}} \exp \biggl( -i \big(\theta_i(\tau) - \theta_j(\tau)\big)\dfrac{\sigma_0+\sigma_3}{2}\biggr) \biggr \} \Psi_j(\tau) \biggr] 
	\end{eqnarray}
	}
	where Nambu spinors $\bar{\Psi}_i(\tau) = (\bar{c}_{i\uparrow}(\tau) , c_{i\downarrow}(\tau))$, and $\sigma_{i = 1,2,3}$'s are the Pauli matrices and $\sigma_0$ is the $2\times2$ identity matrix. This asymmetric gauge transformation for spin-up and spin-down species respects their singlevaluedness which would have been lost for a symmetric gauge transformation between them.

	 Integrating over $\Psi$ and $\bar{\Psi}$, we obtain an effective action $\mathcal{S}_{\rm eff}$ for the bosonic fields $\Delta_i$ and $\theta_i$. Therefore, the partition function becomes,
	 \begin{equation} \label{eqn7}
	 \mathcal{Z} = \int \mathcal{D}[\Delta,\theta]\, e^{-\mathcal{S}_{\rm eff}} 
     \end{equation}
	  with
	\begin{equation}\label{eqn8}
		\mathcal{S}_{\rm eff}  = \int_{0}^{\beta} d\tau
		 \mathlarger{\sum_i} \biggl(\dfrac{|\varphi_i(\tau)|^2}{U}+\dfrac{\xi^2_i(\tau)}{2U}\biggr) - \mathbf{Tr} \bigl[\ln {\mathcal{O}^{-1}(\tau)} \bigr]
	\end{equation}
	where $\mathcal{O}^{-1} = \mathcal{G}^{-1} - \Sigma$ and the expression for phase-independent inverse Green's function $\mathcal{G}^{-1}$ and exclusive phase dependent self energy $\Sigma$ are given by
	\begin{equation}\label{eqn9}
		\begin{split}
			\mathcal{G}_{ij}^{-1}(\tau) & = \Bigl[ \{ - \sigma_0 \partial_{\tau} + (\tilde{\mu_i} - V_i)\sigma_3 - \Delta_i\sigma_1 \} \delta_{ij} \\ 
			& + \sigma_3\, t \sum_{\bm{\delta}=\pm \hat{x},\pm \hat{y}} \delta_{j,i+\bm{\delta}} \Bigr]
		\end{split}
	\end{equation}
	and 
	\begin{equation}\label{eqn10}
		\Sigma_{ij}(\tau)  =  \bigr \{\dot{\theta}_i(\tau) \delta_{ij}
		 + t\sum_\delta \delta_{j,i+\delta}[1-\cos(\theta_{ij}) -i\sin(\theta_{ij})] \bigr\}\sigma^{\prime}  \\
	\end{equation}
	where $\theta_{ij} = (\theta_{i}(\tau) - \theta_{j}(0))$ and $\sigma^{\prime} = (\sigma_0+\sigma_3)/2 $. Here, $\mathbf{Tr}$ represents trace over both space-time and spin degrees of freedom. We next express $\mathcal{S}_{\rm eff}  = \mathcal{S}_0 + \mathcal{S}_\theta$, where phase-independent and phase-dependent actions are respectively given by  
	\begin{equation}\label{amp_action}
		 \mathcal{S}_0  = \int_{0}^{\beta} d\tau 
		 \mathlarger{\sum_i} \biggl(\dfrac{|\varphi_i(\tau)|^2}{U}+\dfrac{\xi^2_i(\tau)}{2U}\biggr) \\
		  -   \mathbf{Tr}[\ln {\mathcal{G}^{-1}(\tau)}] 
	\end{equation}
	and
	\begin{equation}\label{phase_action}
		\mathcal{S}_{\theta}  = -   \mathbf{Tr}\, \bigl[ \ln (\sigma_0 -\mathcal{G}\Sigma) \bigr] =\sum_{n=1}^\infty \frac{1}{n} \mathbf{Tr} \, \bigl[ (\mathcal{G}\Sigma)^n \bigr] .
	\end{equation}

In the clean limit, there exist \cite{Kopec_1993, Kopec_2002} such modulus-phase representations for the negative-$U$ Hubbard model to study the spatial and temporal fluctuations.

\subsection{\label{subsec:A} Phase Action and Effective XY Model}

We expand $\mathcal{S}_{\theta}$ in Eq.\eqref{phase_action} up to quadratic order (Gaussian or harmonic approximation) to obtain an effective phase action whose form will appear like a nearest-neighbor $XY$-model. In the first order,
\begin{eqnarray}\label{phase_action_1st}
		\mathcal{S}^{(1)}_{\theta} 
		&& = -t \sum_{\langle ij\rangle} \int_0^\beta  d\tau   \hspace{0.2em}\mathbf{tr} \bigl[\mathcal{G}_{ij}(\tau^+ - \tau) \sigma^{\prime} \bigr]  \cos(\theta_{ij}) \nonumber\\
		&& = - \beta \sum_{\langle ij\rangle}J_{ij}^{(1)} \mathbf{S}_i\cdot \mathbf{S}_j
\end{eqnarray}
where $\mathbf{S}_i = (\cos \theta_i,\,\sin \theta_i)$, and $\mathbf{tr}$ represents trace over spin degree of freedom alone and $1^{st}$ order coupling constant or local phase stiffness $J_{ij}^{(1)}$ is given by,
\begin{equation}\label{J_1st}
		J_{ij}^{(1)} = \dfrac{t}{\beta}  \int_0^\beta d\tau \hspace{0.2em} \mathbf{tr} \bigl[\mathcal{G}_{ij}(\tau^+ - \tau) \sigma^{\prime}  \bigr]  = \dfrac{t}{\beta}  \sum_{\omega_n}   \hspace{0.2em}\mathbf{tr} \bigl[\mathcal{G}_{ij}(i\omega_n) \sigma^{\prime}  \bigr]
\end{equation}
written in terms of fermionic Matsubara frequency $\omega_n = (2n+1)\pi T$.

In the second order ($n = 2$), the form of local phase stiffnes $J_{ij}^{(2)}$ is given by (see Appendix~\ref{app:A} for detailed calculations),
\begin{eqnarray}\label{J_2nd}
		J_{ij}^{(2)} && = \dfrac{t^2}{\beta} \sum_{\omega_n} \Bigl \{  \mathbf{tr} \bigl[\mathcal{G}_{ii}(i\omega_n)\sigma^{\prime} \mathcal{G}_{jj}(i\omega_n)\sigma^{\prime}\bigr] \nonumber\\
		&& + \sum_{\langle\bm{\delta \delta^{\prime}} \rangle} \mathbf{tr} \bigl[\mathcal{G}_{i,i+\bm{\delta}}(i\omega_n)\sigma^{\prime}\mathcal{G}_{j+\bm{\delta^{\prime}},j}(i\omega_n)\sigma^{\prime}\bigr] \Bigr\}		  
\end{eqnarray}
where $\bm{\delta},\bm{\delta^\prime}$ are nearest-neighbor to $i^{th}$ and $j^{th}$ site respectively and angular bracket represents nearest-neighbor condition. \\
Henceforth, the phase action can be written as
\begin{equation} \label{phase_H}
		\mathcal{S}_{\theta} = - \beta \sum_{\langle ij\rangle} \bigl(J_{ij}^{(1)} + J_{ij}^{(2)} \bigr) \mathbf{S}_i\cdot \mathbf{S}_j = \beta H_{XY}
\end{equation}
where 
\begin{equation}
 H_{XY} = - \sum_{\langle ij\rangle} (J_{ij}^{(1)} + J_{ij}^{(2)}) \mathbf{S}_i\cdot \mathbf{S}_j = - \sum_{\langle ij\rangle} J_{ij}\mathbf{S}_i\cdot \mathbf{S}_j 
 \label{Hamiltonian_xy}
\end{equation}
is the XY-model Hamiltonian.

\begin{figure}[t!]
	\centering
	\begin{minipage}[b]{0.19\textwidth}
		\centering
		\subfloat[]{\begin{tikzpicture}[scale=0.7]
				\foreach \x/\y/\name in {0/0/, 1/0/, 2/0/, 3/0/, 0/1/, 1/1/, 2/1/, 3/1/, 0/2/, 1/2/, 2/2/, 3/2/}
				\node (\name) at (\x,\y) [circle,fill,inner sep=1.2pt,label=below:$\name$] {};
				
				\node[draw, circle, fill=red, inner sep=1.5pt, label=right:\small{$j$}] (j) at (2,1) {};
				\node[draw, circle, fill=red, inner sep=1.5pt, label=left:\small{$i$}] (i) at (1,1) {};
				
				\begin{feynman}
					\vertex (i) at (1,1) ;
					\vertex (j) at (2,1) ;
					\diagram*{
						(j)  -- [fermion,line width=1.2pt,edge label={\small{$\mathcal{G}_{ij}$}},out=180,in=0] (i)
					};
				\end{feynman}
			\end{tikzpicture}}
	\end{minipage}\hfill
	\begin{minipage}[b]{0.19\textwidth}
		\centering
		\subfloat[]{\begin{tikzpicture}[scale=0.7]
				\foreach \x/\y/\name in {0/0/, 1/0/, 2/0/, 3/0/, 0/1/, 1/1/, 2/1/, 3/1/, 0/2/, 1/2/, 2/2/, 3/2/}
				\node (\name) at (\x,\y) [circle,fill,inner sep=1.2pt,label=below:$\name$] {};
				
				\node[draw, circle, fill=red, inner sep=1.4pt, label=right:\small{$j$}] (j) at (2,1) {};
				\node[draw, circle, fill=red, inner sep=1.4pt, label=left:\small{$i$}] (i) at (1,1) {};
				
				\begin{feynman}
					\vertex (i) at (1,1) ;
					\vertex[above=0.6cm of i](t);
					\vertex[above left=0.3cm of i] {\small{$\mathcal{G}_{ii}$}};
					\diagram*{
						(i)  -- [line width=1.2pt,out=135,in=180] (t) --[fermion,line width=1.2pt,out=0,in=45] (i)
					};
				\end{feynman}
				\begin{feynman}
					\vertex (j) at (2,1) ;
					\vertex[above=0.6cm of j](t);
					\vertex[above right=0.3cm of j] {\small{$\mathcal{G}_{jj}$}};
					\diagram*{
						(j)  -- [line width=1.2pt,out=135,in=180] (t) --[fermion,line width=1.2pt,out=0,in=45] (j)
					};
				\end{feynman}
		\end{tikzpicture}}
	\end{minipage}
	\begin{minipage}[b]{0.19\textwidth}
		\centering
		\subfloat[]{\begin{tikzpicture}[scale=0.7]
				\foreach \x/\y/\name in {0/0/, 1/0/, 2/0/, 3/0/, 0/1/, 1/1/, 2/1/, 3/1/, 0/2/, 1/2/, 2/2/, 3/2/}
				\node (\name) at (\x,\y) [circle,fill,inner sep=1.2pt,label=below:$\name$] {};
				
				\node[draw, circle, fill=red, inner sep=1.5pt, label=right:\small{$j$}] (j) at (2,1) {};
				\node[draw, circle, fill=red, inner sep=1.5pt, label=left:\small{$i$}] (i) at (1,1) {};
				
				\begin{feynman}
					\vertex (i) at (1,1) ;
					\vertex (j) at (2,1) ;
					\diagram*{
						(j)  -- [fermion,line width=1.2pt,edge label'={\small{$\mathcal{G}_{ij}$}},out=120,in=60] (i)
					};
				\end{feynman}
				\begin{feynman}
					\vertex (i) at (1,1) ;
					\vertex (j) at (2,1) ;
					\diagram*{
						(j)  -- [fermion,line width=1.2pt,edge label={\small{$\mathcal{G}_{ij}$}}, out=240,in=300] (i)
					};
				\end{feynman}
		\end{tikzpicture}}
	\end{minipage}\hfill
	\begin{minipage}[b]{0.19\textwidth}
		\centering
		\subfloat[]{\begin{tikzpicture}[scale=0.7]
				\foreach \x/\y/\name in {0/0/, 1/0/, 2/0/, 3/0/, 0/1/, 1/1/, 2/1/, 3/1/, 0/2/, 1/2/, 2/2/, 3/2/}
				\node (\name) at (\x,\y) [circle,fill,inner sep=1.2pt,label=below:$\name$] {};
				
				\node[draw, circle, fill=red, inner sep=1.4pt, label=right:\small{$j$}] (j) at (2,1) {};
				\node[draw, circle, fill=red, inner sep=1.4pt, label=left:\small{$i$}] (i) at (1,1) {};
				\node[draw, circle, fill=green, inner sep=1.4pt, label=above:\small{$i+\bm{\delta}$}] (l) at (1,2) {};
				\node[draw, circle, fill=blue, inner sep=1.4pt, label=left:\small{$i+\bm{\delta}$}]  at (0,1) {};
				\node[draw, circle, fill=yellow, inner sep=1.4pt, label=below:\small{$i+\bm{\delta}$}]  at (1,0) {};
				
				\begin{feynman}
					\vertex (i) at (1,1) ;
					\vertex (j) at (2,1) ;
					\vertex (l) at (1,2) ;
					\diagram*{
						(l)  -- [fermion,line width=1.2pt, edge label'=\small{$\mathcal{G}_{i,i+\bm{\delta}}$}, out=90,in=270] (i)
					};
					\diagram*{
						(j)  -- [fermion,line width=1.2pt, edge label=\small{$\mathcal{G}_{ij}$}, out=180,in=0] (i)
					};
				\end{feynman}
		\end{tikzpicture}}
	\end{minipage}
	\begin{minipage}[b]{0.19\textwidth}
		\centering
		\subfloat[]{\begin{tikzpicture}[scale=0.7]
				\foreach \x/\y/\name in {0/0/, 1/0/, 2/0/, 3/0/, 0/1/, 1/1/, 2/1/, 3/1/, 0/2/, 1/2/, 2/2/, 3/2/}
				\node (\name) at (\x,\y) [circle,fill,inner sep=1.2pt,label=below:$\name$] {};
				
				\node[draw, circle, fill=red, inner sep=1.4pt, label=right:\small{$j$}] (j) at (2,1) {};
				\node[draw, circle, fill=red, inner sep=1.4pt, label=left:\small{$i$}] (i) at (1,1) {};
				\node[draw, circle, fill=green, inner sep=1.4pt, label=above:\small{$j+\bm{\delta^\prime}$}] (k) at (2,2) {};
				\node[draw, circle, fill=blue, inner sep=1.4pt, label=right:\small{$j+\bm{\delta^\prime}$}]  at (3,1) {};
				\node[draw, circle, fill=yellow, inner sep=1.4pt, label=below:\small{$j+\bm{\delta^\prime}$}]  at (2,0) {};
				
				\begin{feynman}
					\vertex (i) at (1,1) ;
					\vertex (j) at (2,1) ;
					\vertex (k) at (2,2) ;
					\diagram*{
						(j)  -- [fermion,line width=1.2pt, edge label=\small{$\mathcal{G}_{ij}$}, out=180,in=0] (i)
					};
					\diagram*{
						(j)  -- [fermion,line width=1.2pt, edge label'=\small{$\mathcal{G}_{j+\bm{\delta^\prime},j}$}, out=90,in=270] (k)
					};
				\end{feynman}
		\end{tikzpicture}}
	\end{minipage}\hfill
	\begin{minipage}[b]{0.19\textwidth}
		\centering
		\subfloat[]{\begin{tikzpicture}[scale=0.7]
				\foreach \x/\y/\name in {0/0/, 1/0/, 2/0/, 3/0/, 0/1/, 1/1/, 2/1/, 3/1/, 0/2/, 1/2/, 2/2/, 3/2/}
				\node (\name) at (\x,\y) [circle,fill,inner sep=1.2pt,label=below:$\name$] {};
				
				\node[draw, circle, fill=red, inner sep=1.4pt, label=right:\small{$j$}] (j) at (2,1) {};
				\node[draw, circle, fill=red, inner sep=1.4pt, label=left:\small{$i$}] (i) at (1,1) {};
				\node[draw, circle, fill=green, inner sep=1.4pt, label=above:\small{$j+\bm{\delta^{\prime}}$}] (k) at (2,2) {};
				\node[draw, circle, fill=blue, inner sep=1.4pt, label=below:\small{$j+\bm{\delta^{\prime}}$}] (k) at (2,0) {};
				\node[draw, circle, fill=green, inner sep=1.4pt, label=above:\small{$i+\bm{\delta}$}] (l) at (1,2) {};
				\node[draw, circle, fill=blue, inner sep=1.4pt, label=below:\small{$i+\bm{\delta}$}] (l) at (1,0) {};
				
				\begin{feynman}
					\vertex (i) at (1,1) ;
					\vertex (j) at (2,1) ;
					\vertex (l) at (1,2) ;
					\vertex (k) at (2,2) ;
					\diagram*{
						(l)  -- [fermion,line width=1.2pt, edge label'=\small{$\mathcal{G}_{i,i+\bm{\delta}}$}, out=90,in=270] (i)
					};
					\diagram*{
						(j)  -- [fermion,line width=1.2pt, edge label'=\small{$\mathcal{G}_{j+\bm{\delta^{\prime}},j}$}, out=270,in=90] (k)
					};
				\end{feynman}
		\end{tikzpicture}}
	\end{minipage}
	\label{fig:Jij_2}
	\caption{(color online) Diagrammatic representations of possible contributions in the expressions of $ J_{ij}^{(1)}$ (a) and   $J_{ij}^{(2)}$ (b)--(f) due to single and double hopping respectively between the $i^{th}$ and $j^{th}$ sites (red dots). Lattice points are represented by dots and a line between the $i^{th}$ and $j^{th}$ sites represent ${\cal G}_{ij}$ with an arrow headed towards $i^{th}$ site. (a) Direct coupling between the $i^{th}$ and $j^{th}$ sites, (b) onsite contributions without any involvement of intersite hopping, (c) contribution for double-hopping between the $i^{th}$ and $j^{th}$ sites, (d) and (e) contributions for hoppings involving the $i^{th}$ and $j^{th}$ sites and one of these sites with three other neighboring sites (green, blue, and yellow dots), and (f) the combination of one hopping from the  $i^{th}$ and the other from the  $j^{th}$ site such that the end sites become nearest neighbor (no hopping is involved between the $i^{th}$ and $j^{th}$ sites).}
\end{figure}
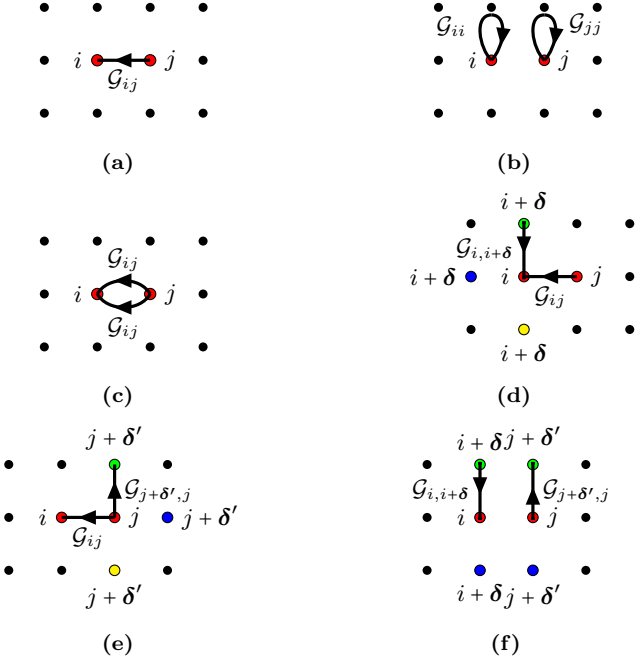

Using Nambu Green's function (see Appendix~\ref{app:B} for detailed calculations) and performing Matsubara sum with the identity 
\begin{equation}\label{eqn18}
	\frac{1}{\beta}\sum_{\omega_n} \frac{1}{i\omega_n -E_m} = \frac{1}{e^{\beta E_m}+1}\end{equation}
followed by taking $\beta \to \infty$ limit for
$T = 0$, we find local phase stiffness in the harmonic approximation of phase fluctuations as    
\begin{eqnarray} 
	J_{ij}^{(1)} &=& t \sum_{E_p>0} V_p^{ij}  \label{eqn19} \\
	J_{ij}^{(2)} &=& - t^2 \mathlarger{\sum_{E_{p,q}>0}} \hspace{0.1em}\dfrac{1}{E_p + E_q} \biggl\{  \Bigl( U_p^{ii}V_q^{jj} +V_p^{ii}U_q^{jj} \Bigr) \nonumber \\
&&	+  \sum_{\langle\bm{\delta \delta^{\prime}} \rangle} \Bigl( U_p^{i,i+\bm{\delta}}V_q^{j+\bm{\delta^{\prime}},j} +V_p^{i,i+\bm{\delta}}U_q^{j+\bm{\delta^{\prime}},j} \Bigr)\biggr\}\label{eqn20}
\end{eqnarray}

with \( U_p^{i,i+\bm{\delta}} = u_p^{i}u_p^{i+\bm{\delta}\star} \) and \( V_p^{i,i+\bm{\delta}} = v_p^{i\star}v_p^{i+\bm{\delta}} \), where \( [u^i_p, v^i_p] \) denotes the eigenvector at site index \( i \) corresponding to the eigenvalue \( E_p \) of the BdG Hamiltonian (see Appendix~\ref{app:B}).

\subsection{\label{subsec:C}Phenomenology with XY Model of Random Coupling}

The XY-model Hamiltonian in Eq.~\eqref{Hamiltonian_xy} can be written in terms of angular variables as
\begin{equation}\label{H_xy}
	H_{XY} = -\sum_{\langle ij\rangle} J_{ij} \hspace*{0.1em} \cos(\theta_{i} - \theta_{j} )
\end{equation}
 $\mathbf{S}_i = (\cos{\theta_{i}},\sin{\theta_i})$.
 We show below (section III) that $J_{ij}$ is random with both positive and negative values with bimodal distributions for higher strength of random potential $V$.

In linear response theory, the superfluid stiffness characterizes the response of a superconducting system to a transverse gauge field \cite{Fetter,Scalapino_1993}, $\vec{\mathcal{A}}$, which can be minimally coupled to the superconducting phase through a local phase twist induced by the gauge field \cite{Maccari_2019}. This coupling modifies XY-model Hamiltonian of the square lattice to

\begin{equation}\label{H_xy_A}
	H_{XY}(\mathcal{A}) = - \mathlarger{\sum}_{i,\delta=\hat{x},\hat{y}} J_{i,i+\delta} \cos \left(\theta_{i} - \theta_{i+\delta} + \int_{\vec{i}}^{\vec{i}+\vec{\delta}} \vec{{\cal A}} \cdot d\vec{
		l} \right)\, .
\end{equation}

For classical XY model Hamiltonian $H_{XY}(\mathcal{A})$, the superfluid stiffness is calculated from the second derivative of the free energy with respect to the applied phase twist, $J_s = \left. \frac{1}{N} \frac{\partial^2 \mathcal{F}(A_x)}{\partial A_x^2} \right|_{A_x = 0}$, where $A_x$ represents the twist along the $x$-direction \cite{Schultka_1994}. Within this framework, the stiffness separates into two distinct contributions: a diamagnetic term, $J_d$, arising from the average energy associated with nearest-neighbor phase correlations, and a paramagnetic term, $J_p$, arising from current fluctuations. The superfluid stiffness is then given by $J_s = J_d - J_p$, with the individual terms defined as \cite{Schultka_1994,Maccari_2019}

\begin{eqnarray}
	 {\large J_d} &=& \frac{1}{N} \overline{\biggl \langle \mathlarger{\sum}_{i,\delta =\hat{x},\hat{y}} J_{i,i+\delta}  \cos(\theta_{i} - \theta_{i+\delta})\biggr \rangle} \label{J_d} \\
	 {\large J_p} &=&  \frac{1}{NT} \overline{\biggl \langle \biggl[\mathlarger{\sum}_{i,\delta =\hat{x},\hat{y}} J_{i,i+\delta}  \sin(\theta_{i} - \theta_{i+\delta})\biggr]^2\biggr \rangle} \label{J_p}
\end{eqnarray}
with $\langle ... \rangle$ and $\overline{(...)}$ being averages over the thermal degrees of freedom and over the random configurations of $J_{ij}$ respectively. Here $N$ is the number of lattice points, and  $T$ is the temperature.

In the clean limit and at zero temperature, the system exhibits a purely diamagnetic response, reflecting perfect flux expulsion. However, at finite temperatures, thermally activated phase fluctuations give rise to time-dependent local phase gradients which induce fluctuating currents. This dynamic effect is captured by the current-current correlation function and constitutes the paramagnetic response, $J_p$.

In the presence of disorder, even at $T = 0$, the breaking of translational symmetry leads to phase gradients, which produce persistent local currents leading to a finite paramagnetic contribution, suppressing the superfluid stiffness. 

The sign of $J_{ij}$ becomes randomly negative at certain sites for any configuration of disorder at higher strength, making it possible to have a glassy phase. We thus calculate the Edwards-Anderson order parameter $q_{_{\rm EA}}$ \cite{Parisi_1983}, defined as
\begin{equation}\label{EA_order}
	q_{_{\rm EA}} = (1/N) \overline{ \bigl[\sum_{i} {\lvert \langle \mathbf{S}_i \rangle \rvert}^2 \bigr] }.
\end{equation}

The phase Hamiltonian $H_{XY}$ in Eq.~\eqref{H_xy} is invariant under a global rotation of phase configuration, which leads to an arbitrary global phase rotation during the MC updating steps. Taking care of the uniform phase rotations, the EA order parameter can be redefined as \cite{Dasgupta_1988}
\begin{equation}\label{EA_redef}
	q_{_{\rm EA}} = \lim_{t\rightarrow \infty}  \overline{ \bigl[ \bigl\langle (1/N)\sum_{i}  \mathbf{S}_i(0) \hspace*{0.2em} \mathcal{R}(\phi) \hspace*{0.2em} \mathbf{S}_i(t) \bigr\rangle \bigr] }
\end{equation}
where $\mathcal{R}(\phi)$ is the general SO(2) rotational matrix and the global rotational angle $\phi$ that maximizes  $(1/N)\sum_{i}  \mathbf{S}_i(0) \hspace*{0.2em} \mathcal{R}(\phi) \hspace*{0.2em} \mathbf{S}_i(t)$ is given by
\begin{equation}\label{rot_angle}
	\tan(\phi) = \sum_{i} \sin(\theta_i(0)-\theta_i(t)) / \sum_{i} \cos(\theta_i(0)-\theta_i(t)).
\end{equation}

	\section{\label{sec:III} Numerical RESULTS}
	\subsection{\label{res:A}Probability Distribution of Random Local Phase Stiffness }	
	
	We solve the BdG eigenvalue equation (see Appendix \ref{app:B}) numerically in a self-consistency manner on a square lattice of $N = 48 \times 48$ sites with periodic boundary conditions. Interaction strength has been taken to be $U = 1.5t$. The average electron density is set to $ \langle n \rangle = 0.875 $, which is sufficiently close to half-filling to capture correlation effects but not so close that the system becomes dominated by strong correlation or charge-ordering phenomena. This choice of density provides a balance, allowing the study of how the disorder impacts the superconducting gap, density of states, and superfluid stiffness, which have been extensively studied previously \cite{Ghoshal_2001, SSM_2013}. We choose white-noise disorder potential ranging from the clean case ($V = 0.0t$) to $V = 3.0t$, and all the calculations are statistically averaged at over \(25\) disorder realizations. Henceforth, we express all the length scales in the unit of lattice constant, and all the energy scales are in the unit of nearest-neighbor hopping integral $t$.
	
\begin{figure}[t!]
		\includegraphics[width=\linewidth]{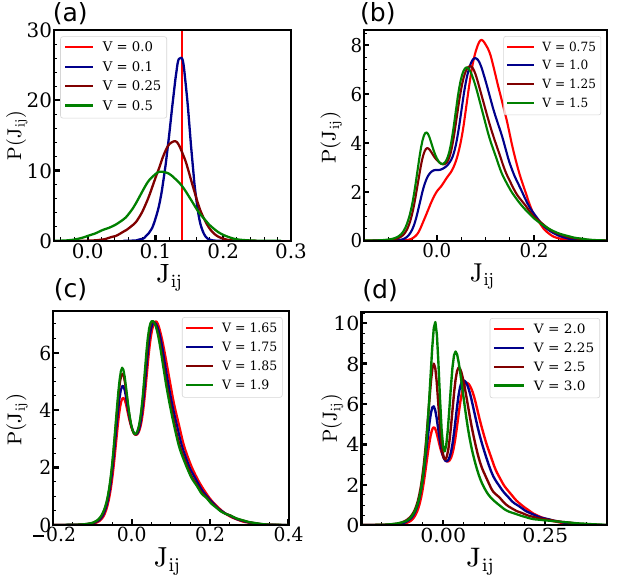}
		\caption{(color online) Probability distribution functions of nearest neighbor coupling $J_{ij}$ for different disorder strengths $V$ in a $48 \times 48$ square lattice. Four panels correspond to different ranges of $V$.}
		\label{fig:J_ij}
	\end{figure}

	Using BdG eigenvalues and eigenvectors in Eqs.~\eqref{eqn19}-\eqref{eqn20}, we compute the bond-dependent local phase stiffness $J_{ij}$ at the zero temperature. Figure~(\ref{fig:J_ij}) shows the distribution \( P(J_{ij}) \) of the local coupling \( J_{ij} \) for different values of disorder strength \( V \). In the clean limit (\( V = 0 \)), the distribution exhibits a sharp line at \( J_0 \approx 0.139 \), Bardeen-Cooper-Schrieffer (BCS) value, indicating uniform coupling throughout the lattice. A small increase of disorder strength ($V=0.1$) makes the distribution slightly broadened with a sharp peak at a value slightly below the BCS value $J_{0}$. At this strength of $V$, the eigenstates of the BdG Hamiltonian are still spatially uniform, resulting in a homogeneous distribution of local couplings. This uniformity across all nearest-neighbor bonds renders the system robust against external phase twists.
	
	\begin{figure*}[t!]
		\includegraphics[width=\linewidth]{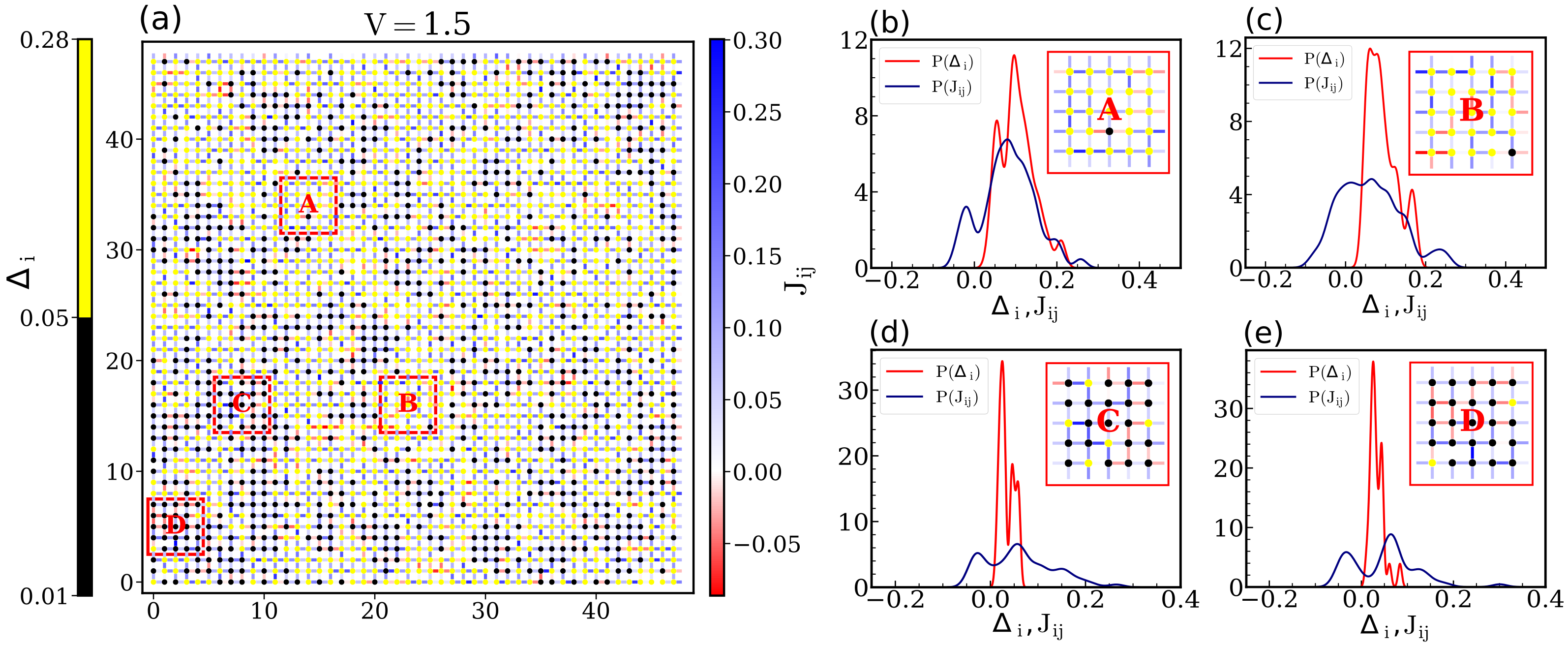}
		\caption{(color online) (a) Color-coded maps of the lattice variable $\Delta_i$ and the nearest-neighbor bond-variable $J_{ij}$ on a $48\times48$ square lattice for a disorder realization with strength $V=1.5$. Lesser (greater) values of $\Delta_i$ are marked with black (yellow) color. (b)--(e) Probability distributions functions $P(\Delta_i)$ and $P(J_{ij})$ for the selected zones A--D in (a) respectively. Insets: Magnified view of the selected zones.}
		\label{fig:J_ij_delta_i_lattice_view}
	\end{figure*}
	
	 As \( V \) increases further (\( V \simeq 0.5 \)), the distributions maintain a peak characterized by an increasing broadening and a decreasing mean value. In this regime, our microscopic calculation of the local coupling distribution is consistent with previous studies \cite{Maccari_2017, Maccari_2019} on the disordered phase-only XY model in superconductors, where Gaussian distributions are assumed in the weak disorder limit. 
	
	For a moderate disorder potential (\( V \simeq 0.75 \)), the left tail of the distribution extends slightly into negative coupling values, making it negatively skewed. The emergence of negative local stiffness adds an intriguing aspect to our theoretical analysis. Notably, at \( V = 1.0 \), a distinct hump begins to form at some negative values of $J_{ij}$, and it is converted into a peak for further increase of $V$. In the strong disorder regime (\( V \gtrsim 1.5 \)), this secondary peak becomes substantial, resulting in a bimodal distribution: one conventional peak at positive $J_{ij}$ and the other peak at negative $J_{ij}$. This distribution, however, is in contrast to a prior study \cite{Maccari_2019} where a two-delta distribution was assumed with the lowest value occurring at $J_{ij} = 0$. The peak at negative $J_{ij}$ raises three important questions: (a) How does local coupling correlate with local pairing amplitude? (b) What is the origin of these negative coupling values? (c) Why does the distribution exhibit a secondary peak at \( J_{ij} < 0 \) for moderate to high disorder strengths?
	
	\begin{figure}[b!]
		\includegraphics[width=\linewidth]{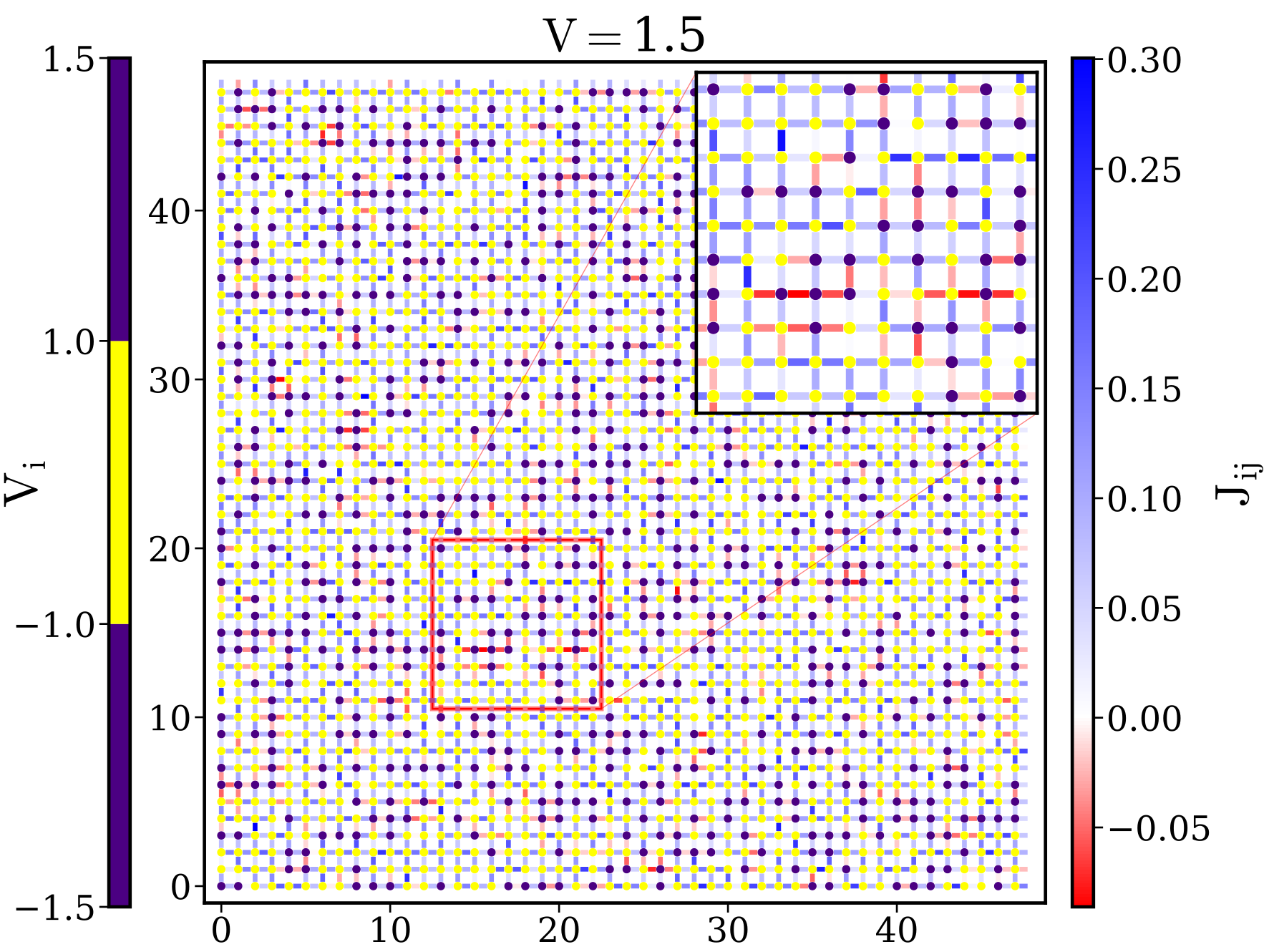}
		\caption{(color online) In a $48\times 48$ square lattice for a realization of disorder with $V=1.5$, onsite disorder potential $V_i$ and bond-variable (between nearest neighbors) $J_{ij}$ are shown with color-coding. $\vert V_i\vert <1\, (>1)$ is represented by yellow (indigo) points. Inset: Magnified view of a selected square-shaped zone with red-colored boundary.}  
		\label{fig:J_ij_V_i_lattice_view}
	\end{figure}
	
	To explore the correlation between local couplings and the local pairing amplitude, we show (Fig.~\ref{fig:J_ij_delta_i_lattice_view}) \( J_{ij} \) across two nearest-neighbor sites \( i \) and \( j \), alongside the local pairing amplitude \( \Delta_i \) on a square lattice for \( V = 1.5 \). The color variation of \( J_{ij} \) smoothly transitions from deep red (representing the most negative values) to deep blue (representing the most positive values). For \( \Delta_i \), two distinct colors are used: black for \( \Delta_i \lesssim 0.05 \) and yellow for \( 0.05 \lesssim \Delta_i \lesssim 0.28 \). We next select four \( 5 \times 5 \) square regions: two (A and B) within the superconducting island (high \( \Delta_i \) region), and two (C and D) in the non-superconducting sea (low \( \Delta_i \) region). The probability distributions \( P(J_{ij}) \) and \( P(\Delta_i) \) for these regions are shown in Figs.~\ref{fig:J_ij_delta_i_lattice_view}(b)--(e). It is evident that \( P(J_{ij}) \) spans from negative to positive values in all four regions, regardless of the \( \Delta_i \) distributions. A zoomed-in view (insets in Figs.~(b)--(e)) further reveals that there is no strict correlation between high \( \Delta_i \) and positive $J_{ij}$ or low \( \Delta_i \) and negative $J_{ij}$. These observations suggest that while the pairing amplitude exhibits clustering at moderate to high disorder strengths, forming superconducting islands within a non-superconducting background, no such clustering is observed in the distribution of local phase stiffness. In other words, no significant correlation is found between the local phase stiffness and the local pairing amplitude.
	
	To further investigate the origin of negative \( J_{ij} \) and its correlation with the local disorder potential \( V_i \), we present (Fig.~\ref{fig:J_ij_V_i_lattice_view}) \( J_{ij} \) across the bond between nearest-neighbor sites \( i \) and \( j \), alongside the onsite disorder potential \( V_i \), on a \( 48 \times 48 \) square lattice. Our analysis reveals that negative coupling predominantly occurs between pairs of nearest-neighbor sites where the disorder potential exhibits deep valleys or high peaks (indigo in the colorbar of $V_i$ in Fig.~\ref{fig:J_ij_V_i_lattice_view}). When itinerant electrons encounter a site with a deep valley, they become trapped in the potential well, resulting in double occupancy at that site. Conversely, sites corresponding to high mountains in the potential energy landscape are effectively inaccessible due to the inability of electrons to hop onto them, leading to zero occupancy. Local pairing amplitude mostly vanishes in this region, making locally insulating region \cite{Ghoshal_2001}. In a homogeneously disordered superconductor, when a doubly occcupied insulating region couples two SC regions, electrons can tunnel through it above the Fermi surface (direct tunnelling) and another one below the Fermi surface (indirect tunnelling). Interference between these processes, particularly when the indirect process dominates the direct one, can result in a negative coupling, as demonstrated by Kivelson and Spivak \cite{Kivelson_1991,Kivelson_1992}. For the case of \(V=1.5\), we find that negative coupling values are predominantly associated with sites having potential \( |V_i| \gtrsim 1.0 \), as shown in Fig.~\ref{fig:J_ij_V_i_lattice_view}. Since the distribution of \( V \) is uniform, increasing the disorder strength leads to more sites with higher potentials, resulting in a secondary peak on the \( J_{ij} < 0 \) side, with larger potential values contributing to the most negative tail of the secondary peak.
	
	As the disorder strength increases, we find that \(J_{ij}\) not only becomes highly inhomogeneous but also develops a mixture of ferromagnetic (positive) and antiferromagnetic (negative) nearest-neighbor couplings. This combination of competing interactions introduces frustration into the phase configurations of the system.

	\subsection{\label{res:B} Superfluid Density and Critical Temperature}

    To determine the temperature-dependent superfluid density and the BKT transition temperature, we perform Monte Carlo simulations of the XY-model Hamiltonian \(H_{XY}\) given in Eq.~\eqref{H_xy} on a \(64 \times 64\) square lattice with periodic boundary conditions. The local couplings \( J_{ij} \) are randomly assigned following the probability distributions \( P(J_{ij}) \) shown in Fig.~\ref{fig:J_ij}. We have employed the Wolff cluster algorithm \cite{Wolff_1989} for faster simulation and reducing critical slowing down effects. The system was annealed from high to low temperatures across $80$ temperature steps. For each temperature, \( 4 \times 10^5 \) Monte Carlo steps are considered, with quantities averaged over the last \( 3 \times 10^5 \) steps after discarding the transient regime observed during the initial \( 1 \times 10^5 \) steps. Finally, disorder averaging is carried out for each disorder level using $25$ independent configurations.
    
    \begin{figure}[t!]
    	\includegraphics[width=\linewidth]{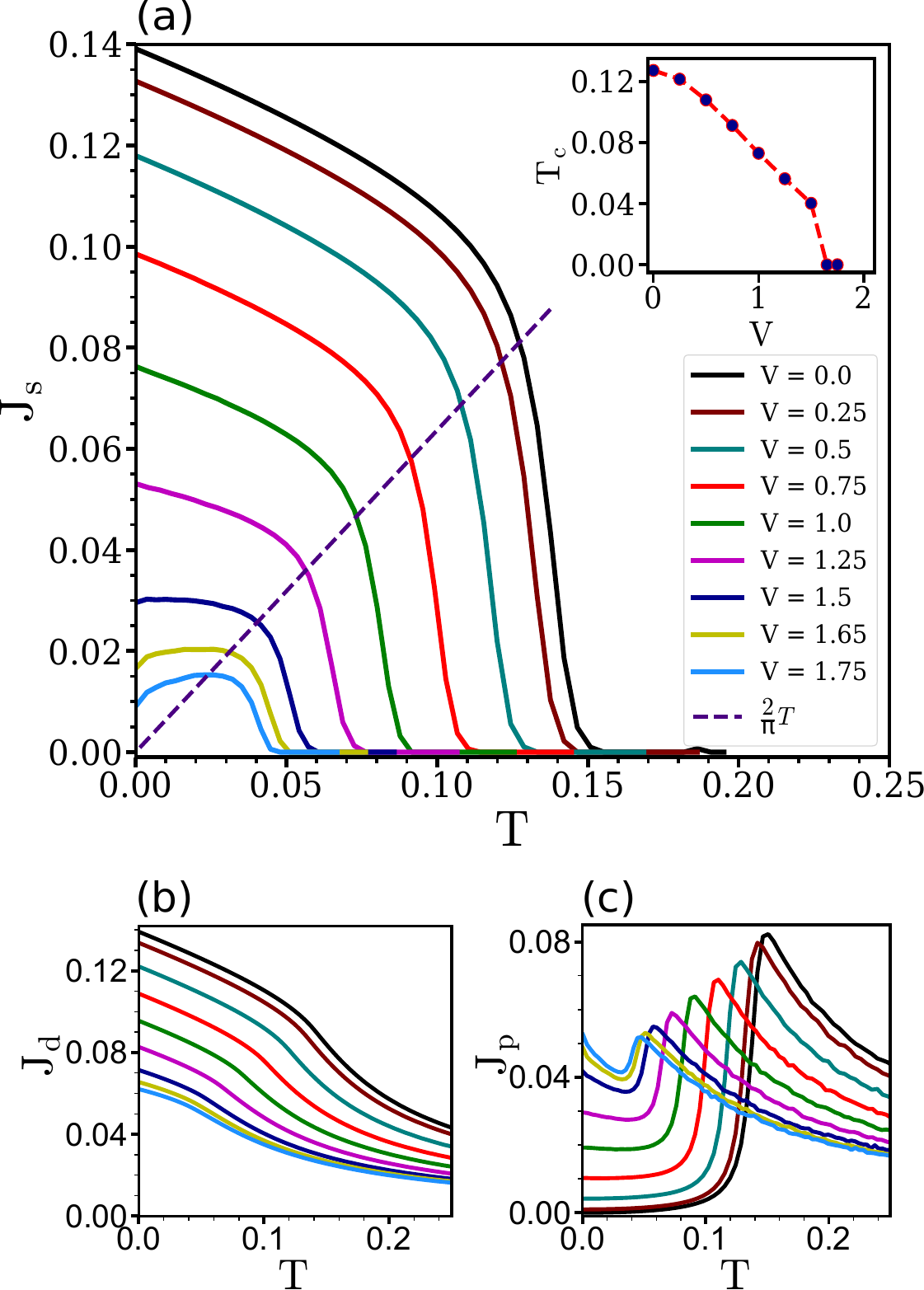}
    	\caption{(color online) Temperature dependence of the superfluid stiffness (a), its diamagnetic contribution (b), and its paramagnetic contribution (c). The dashed lines in (a) correspond to $J_s = (2/\pi)T$ as per the BKT criterion for BKT transition from superfluid state to the normal state. All figures share the same color legend. Inset: Variation of the critical temperature \(T_c\) with disorder strength \(V\).}
    	\label{fig:stiffness}
    \end{figure}
	
	In the clean limit (\( V = 0 \)), Monte Carlo simulations [Fig.~\ref{fig:stiffness} a,b,c] shows that the temperature dependence of the superfluid stiffness up to \( T = 0.08 \) is primarily governed by diamagnetic suppression. At low temperatures, the only significant excitations are thermally induced low-energy longitudinal phase modes, which lead to a reduction in the diamagnetic term $J_d$ in Eq.~\eqref{J_d}. As the temperature increases, vortex-antivortex pairs begin to form and eventually unbind at the BKT transition. This unbinding mainly affects the paramagnetic contribution, which captures long-range current correlations, whereas the diamagnetic contribution remains predominantly local. Consequently, \( J_p \) in Eq.~\eqref{J_p} sharply rises (Fig.~\ref{fig:stiffness}c) as the system approaches \( T_c \), which is estimated from the intersection point of the \(\frac{2}{\pi}T\) and \(J_s(T)\) curves, leading to a rapid drop in the superfluid stiffness — a hallmark of the universal BKT jump \cite{Berezinsky_1972}.
	
	In the low disorder regime ($V \lesssim 0.5$), where the \( J_{ij} \) distribution is single-peaked, and in the moderate disorder regime (\( 0.5 \lesssim V \lesssim 1.0 \)), the temperature dependence of \( J_d \) and \( J_p \) remains similar to the clean case. Consequently, the qualitative nature of the stiffness curve is preserved. In this regime, the jump in \( J_s(T) \) occurs at the intersection with the universal \(\frac{2}{\pi}T\) line, maintaining the universal relation \( J_s(T_c) = \frac{2}{\pi}T_c \), consistent with the Harris criterion \cite{Harris_1974}. As a result, the length scale set by disorder remains negligible for the ordering process at criticality. The zero-temperature superfluid stiffness, \( J_s(T \approx 0) \), and the BKT critical temperature, \( T_c \), both decrease gradually with increasing disorder strength. The variation of \( T_c \) with \( V \) is shown in the inset of Fig.~\ref{fig:stiffness}(a).
	
	As the disorder strength increases (\( V = 1.25 - 1.5 \)), the low-temperature suppression of the superfluid stiffness becomes notably flatter, as shown in Fig.~\ref{fig:stiffness}(a). Furthermore, the stiffness no longer exhibits a sharp drop near the phase transition, and the characteristic BKT signature becomes less pronounced in this disorder regime. This low-temperature flattening is attributed entirely to the paramagnetic contribution, \( J_p \). From Fig.~\ref{fig:stiffness}(c), it is evident that \( J_p \) decreases with increasing temperature, with a slope that becomes steeper as the disorder strength increases. This suggests that at low temperatures thermal effects arising from longitudinal phase fluctuations have a more pronounced impact on the paramagnetic term than on the diamagnetic one. This is presumably due to the long-range character of the correlations probed by the current-current response function compared to the local nature of the diamagnetic response \cite{Maccari_2019}.
	
	For \( V \geq 1.65 \), the low-temperature slope of the paramagnetic response \( J_p \) exceeds that of the diamagnetic term \( J_d \), leading to a non-trivial temperature dependence of the superfluid stiffness \( J_s(T) \). Notably, \( J_s(T) \) initially increases with temperature, reaches a peak, and subsequently decreases, ultimately vanishing at the critical temperature. This non-monotonic behavior in the low-temperature regime under strong disorder suggests the emergence of a qualitatively distinct phase. Specifically, it indicates that the system may transition into a non-superconducting state beyond a critical disorder strength. Although this non-monotonicity may not directly manifest in experimental observations, owing to the loss of global phase coherence between superconducting islands, we interpret its unusual nature as indicative of the vanishing of the superconducting transition temperature (\( T_c \approx 0\) ) at \( V = 1.65 \). This assumption provides a plausible basis for identifying the emergence of a disorder-driven quantum phase transition, in agreement with quantum Monte Carlo results \cite{Ghoshal_2001, Nandini_2011}.
	
	Given the unusual thermodynamic behavior observed, we posit the possible emergence of a glassy phase at low temperatures. Prior work has suggested that a sufficiently large fraction of antiferromagnetic (negative) couplings can introduce frustration in the system, potentially driving it into a superconducting glass or insulating glassy phase \cite{Kivelson_1991, Kivelson_1992}. To explore this possibility, we compute the Edwards-Anderson order parameter within our Monte Carlo simulations as a diagnostic for glassy behavior.
	
\subsection{Edwards-Anderson Order Parameter}

	\begin{figure}[t!]
		\includegraphics[width=\linewidth]{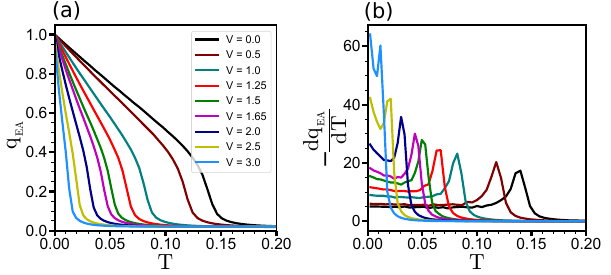}
		\caption{(color online) (a) The Edwards-Anderson order parameter $q_{_{\rm EA}}$ and (b) its temperature derivative $-dq_{_{\rm EA}}/dT$ as functions of temperature $T$ for various disorder strengths $V$. The nature of the curves clearly shows that low-temperature behaviors for $V \lesssim 1.5$ and higher correspond to two different phases. Both figures share the same color legend.}
		\label{fig:EA_order}
	\end{figure}
	
	To investigate the possibility of a glassy transition, we calculate the Edwards-Anderson (EA) order parameter using Eq.~\eqref{EA_redef}, as shown in Fig.~\ref{fig:EA_order}. To determine the EA order parameter at a specific temperature, we performed time averaging of $q_{_{\rm EA}}(t,T)$ using configuration states at time lags that are integer multiples of $1000$ MC steps, while considering global rotations for each pair calculation based on Eq.~\eqref{rot_angle}. Fig.~\ref{fig:EA_order} clearly demonstrates that the EA order parameter behaves distinctly in the presence of strong disorder, exhibiting a much sharper decline with increasing temperature than in the clean or weakly disordered cases. To quantify this behavior, we analyze the plot of the temperature derivative of $q_{_{\rm EA}}(T)$ with $T$, which highlights a pronounced change in the low-temperature regime for disorder strengths $V > 1.5$. This indicates a qualitative shift in the system's response, distinct from that observed at lower disorder levels. The nature of EA order parameter curves mostly agrees with earlier studies \cite{Dasgupta_1988} of the prediction of glassy phases in superconducting systems, although in a slightly different context.  
	
	Our Monte Carlo simulation suggests that a quantum phase transition from a superconducting to a non-superconducting state occurs within the range \( 1.65 \lesssim V \lesssim 1.75 \). Based on our analysis of the superfluid stiffness and the Edwards-Anderson order parameter, this non-superconducting phase may be considered as $phase$-$glass$.
	
	\section{\label{sec:IV} CONCLUSION}
	 In summary, we have analyzed the impact of disorder-induced phase fluctuations on the local phase stiffness in a conventional 2D superconductor. Using the path integral approach, we decomposed the effective action into two components: the mean-field term, treated using the standard BdG formalism, and the phase fluctuation term. The latter was expanded within the harmonic approximation and mapped onto an XY phase-only model Hamiltonian, with the effects of disorder incorporated into the local coupling parameters. Our findings show that at moderate to high disorder strengths, while the pairing amplitude forms superconducting islands within a non-superconducting background, no such clustering is observed in the local couplings, which remain randomly distributed throughout the lattice. At low disorder, phase couplings follow a single-peak distribution, with the width and mean-value quantifying disorder strength. As disorder increases, some local phase couplings become negative, leading to a bimodal distribution with a secondary peak at negative \( J_{ij} \). Additionally, sufficiently negative couplings can drive the system into a phase-frustrated state. These results open new avenues for exploring phase dynamics near the superconductor-to-insulator transition.
	 
	Our Monte Carlo simulation results reveal that, in the presence of weak disorder, the low-temperature behavior of the superfluid stiffness is predominantly governed by the diamagnetic response, while the paramagnetic response becomes more influential near the critical temperature and makes a rapid downturn of stiffness, which the universal BKT jump in simulations on a finite-size system. In contrast, at moderate to strong disorder, the presence of negative local couplings significantly alters the low-temperature behavior of the stiffness, which becomes predominantly controlled by the paramagnetic response. The low-temperature variation of the stiffness exhibits substantial flattening, and the universal jump near \( T_c \) is smeared out due to the inhomogeneous coupling parameters with a mixture of positive and negative signs. In the presence of strong disorder, the bimodal distribution of coupling constants gives rise to anomalous low-temperature behavior in the superfluid stiffness, indicative of underlying fluctuation-driven mechanisms rather than conventional thermal suppression. In this regime, the temperature dependence of the Edwards-Anderson order parameter indicates the possible emergence of $phase$-$glass$ state at low temperatures. Our study, based on mean-field calculations using the BdG approach and incorporating thermal phase fluctuations, does not account for quantum fluctuations. We believe incorporating quantum fluctuations within a phase-only model would provide a more comprehensive understanding of the nature of this glassy state. Indeed, there are suggestions of an anomalous metallic phase at $T \rightarrow 0$ with increasing disorder (or other parameters) \cite{Kapitulnik_2019}, dubbed “failed superconductors”, where the electrical conduction is governed by bosonic (Cooper pair) quantum fluctuations. Inhomogeneities in some superconductors, however, may induce subdominant odd-frequency pairing correlations, leading to a paramagnetic response\cite{Asano_2014}.
	
	The negative phase couplings at several bonds would likely provide nontrivial phenomena in disordered superconductors. For example, by introducing an electromagnetic field in the phase-only Hamiltonian, it would be interesting to study its effect on the pinning of the magnetic vortices.
		
	\section*{\label{sec:VI} ACKNOWLEDGMENTS}
	S.B. thanks Ravi Kiran for computation-related discussions and acknowledges the Ministry of Education, Govt. of India for a research fellowship. A.T. thanks Urmimala Dey for discussions in the early stage. S.S.M. acknowledges Sanjib Ghosh for discussions and helping in obtaining preliminary numerical results at an early stage.
	
	\appendix
	
	\section{\label{app:A} SECOND ORDER LOCAL PHASE STIFFNESS $\mathbf{J^{(2)}_{ij}}$}
	The phase-only part of the action is,
	\begin{equation}\label{A1}
		\tag{A1}
		\mathcal{S}_{\theta} = \sum_{n=1}^\infty \frac{1}{n} \mathbf{Tr} \, \bigl[ (\mathcal{G}\Sigma)^n \bigr] .
	\end{equation}
	Now, the second-order phase action ($n=2$) can be written as follows,
	\begin{widetext}
	\begin{equation}\label{A2}
		\tag{A2}
		\mathcal{S}^{(2)}_{\theta}  
		\approx \dfrac{t^2}{2} \int_0^\beta d\tau \int_0^\beta d\tau^\prime \hspace{0.2em}\sum_{\langle jk \rangle} \sum_{\langle li \rangle}  \{(1- \mathbf{S}_j\cdot \mathbf{S}_k) -i (\mathbf{S}_j \times \mathbf{S}_k)\cdot \hat{z} \} \{(1- \mathbf{S}_l\cdot \mathbf{S}_i) -i (\mathbf{S}_l \times \mathbf{S}_i)\cdot \hat{z} \} \mathbf{tr} \bigl[\mathcal{G}_{ij}(\tau-\tau^\prime) \sigma^{\prime} \mathcal{G}_{kl}(\tau^\prime-\tau) \sigma^{\prime} \bigr] .
	\end{equation}
	\end{widetext}
	Using the identity
	\begin{equation}\label{A3}
		\tag{A3}
		(\mathbf{S}_j \times \mathbf{S}_k)\cdot \hat{z} \,  (\mathbf{S}_l \times \mathbf{S}_i)\cdot \hat{z}  = (\mathbf{S}_j\cdot \mathbf{S}_l)(\mathbf{S_k}\cdot \mathbf{S}_i) - (\mathbf{S}_j\cdot \mathbf{S}_i)(\mathbf{S}_k\cdot \mathbf{S}_l)
	\end{equation}
	we find
	\begin{equation}\label{A4}
		\tag{A4}
		\begin{split}
			&	\{(1- \mathbf{S}_j\cdot \mathbf{S}_k) -i (\mathbf{S}_j \times \mathbf{S}_k)\cdot \hat{z} \} 
			\{(1- \mathbf{S}_l\cdot \mathbf{S}_i) -i (\mathbf{S}_l \times \mathbf{S}_i)\cdot \hat{z} \} \\
			&\approx  - \mathbf{S}_j\cdot \mathbf{S}_k - \mathbf{S}_l\cdot \mathbf{S}_i + (\mathbf{S}_j\cdot \mathbf{S}_k)(\mathbf{S}_l\cdot \mathbf{S}_i) - (\mathbf{S}_j\cdot \mathbf{S}_l)(\mathbf{S}_k\cdot \mathbf{S}_i) \\
			& + (\mathbf{S}_j\cdot \mathbf{S}_i)(\mathbf{S}_k\cdot \mathbf{S}_l)
		\end{split}
	\end{equation}
	retaining only the terms which can contribute to nearest-neighbor $XY$ coupling $\mathbf{S}_i\cdot \mathbf{S}_j$. However, third, fourth and fifth terms in Eq.~\eqref{A4} cannot make such contributions because of the pair of nearest-neighbors conditions $\langle j k\rangle$ and $\langle l i\rangle$ in Eq.~\eqref{A2}. The first two terms in Eq.~\eqref{A4} will make equal contributions in Eq.~\eqref{A2}. 
	
	After substituting $k$ by $i$ and $l$ by $j$ in Eq.~\eqref{A2}, we find
	\begin{equation} \label{A5}
		\tag{A5}
		\mathcal{S}^{(2)}_{\theta} = - \beta \sum_{\langle ij\rangle}J_{ij}^{(2)} \mathbf{S}_i\cdot \mathbf{S}_j
	\end{equation}
	where
	\begin{equation}\label{A6}
		\tag{A6}
		\begin{split}
			J_{ij}^{(2)} & \approx  \dfrac{t^2}{\beta} \int_0^\beta d\tau \int_0^\beta d\tau^\prime \hspace{0.2em} \Bigl \{ \mathbf{tr} \bigl[\mathcal{G}_{ii}(\tau-\tau^\prime)\sigma^{\prime} \mathcal{G}_{jj}(\tau^\prime-\tau) \sigma^{\prime} \bigr] \\
			& \hspace*{0.3cm}+ \sum_{\langle\bm{\delta \delta^{\prime}} \rangle} \mathbf{tr} \bigl[\mathcal{G}_{i,i+\bm{\delta}}(\tau-\tau^\prime)\sigma^{\prime} \mathcal{G}_{j+\bm{\delta^{\prime}},j}(\tau^\prime-\tau) \sigma^{\prime} \bigr] \Bigr \} \\
			& = \dfrac{t^2}{\beta} \sum_{\omega_n} \Bigl \{  \mathbf{tr} \bigl[\mathcal{G}_{ii}(i\omega_n)\sigma^{\prime} \mathcal{G}_{jj}(i\omega_n)\sigma^{\prime}\bigr] \\
			& \hspace*{0.3cm}+ \sum_{\langle\bm{\delta \delta^{\prime}} \rangle} \mathbf{tr} \bigl[\mathcal{G}_{i,i+\bm{\delta}}(i\omega_n)\sigma^{\prime}\mathcal{G}_{j+\bm{\delta^{\prime}},j}(i\omega_n)\sigma^{\prime}\bigr] \Bigr\}		  
		\end{split}
	\end{equation}
	we have redefined $l = i + \bm{\delta}$ and $k = j + \bm{\delta^\prime}$ with $ \bm{\delta},\bm{\delta^\prime}$ ($\pm \hat{x},\pm \hat{y}$ for square lattice) are nearest-neighbors to $i^{th}$ and $j^{th}$ sites respectively and angular bracket represents nearest-neighbor condition.
	
	\section{\label{app:B} NAMBU GREEN'S FUNCTION IN TERMS OF BdG EIGENVECTORS AND ITS CONNECTION WITH THE SADDLE POINT }
	
	Finding the connection between the saddle point of the fermionic action and BdG theory \cite{Samanta_2020}, it is noteworthy that in the fermionic Matsubara frequency domain $\omega_n = (2n+1)\pi T$, the saddle-point inverse Green’s function $\mathcal{G}^{-1}(\omega_n)$ is related to the BdG Hamiltonian $\mathbf{H}_{BdG}$ through
	\begin{equation}\label{B1}
		\tag{B1}
		\mathcal{G}^{-1}(\omega_n) = i \omega_n \mathbb{I} - \mathbf{H}_{BdG}
	\end{equation}
	where the matrix elements of the well-known BdG mean-field Hamiltonian $\mathbf{H}_{BdG}$ is given by
	\begin{equation}\label{B2}
		\tag{B2}
		\mathbf{H}_{BdG}^{ij} = \sigma_3 t \, \delta_{j,i+\bm{\delta}} + \left[ (V_i-\tilde{\mu}_i )\sigma_3 + \Delta_i \sigma_1 \right] \delta_{ij} .
	\end{equation}
	Let \( \Gamma \) represent the unitary transformation that diagonalizes the BdG Hamiltonian. The eigenvalues of the BdG Hamiltonian are denoted as \( E_m \), with the corresponding eigenvectors expressed as \( [u^i_m, v^i_m] \), where \( i \) denotes the site index. Then,
	\begin{equation}\label{B3}
		\tag{B3}
		\Gamma^\dagger \mathbf{H}_{BdG} \Gamma = E_m
	\end{equation}
	where the matrix \( \Gamma \) is constructed from the eigenvectors as columns, \( \Gamma_{im} = u^i_m \) for \( i < N \) and \( \Gamma_{im} = v^i_m \) for \( i \geq N \), with \( N \) representing the total number of lattice sites. The particle-hole symmetry of the BdG Hamiltonian ensures that the eigenvalues occur in pairs \( (E_m, -E_m) \). If \( [u^i_m, v^i_m] \) is the eigenvector corresponding to \( E_m \), then the eigenvector associated with \( -E_m \) is given by \( [-v^{i\star}_m, u^{i\star}_m] \).
	
	The saddle-point action $S_0$ reproduces BdG mean-field theory with the saddle-point equations $\delta S_0/\delta\Delta_i = 0$ and  $\delta S_0/\delta\xi_i = 0$, yields the BdG self-consistency equations at $T = 0$,
	\begin{align}
		\Delta(\mathbf{r_i}) &= U\sum_{E_m > 0} u^i_m v^{i\star}_m  
		\tag{B4} \label{B4} \\
		\xi_i &= U\sum_{E_m > 0} |v^i_m|^2  
		\tag{B5} \label{B5} \\
		\langle n_e \rangle &= \frac{2}{N}\sum_{E_m > 0,i} |v^i_m|^2  
		\tag{B6} \label{B6}
	\end{align}
	
	where $\langle n_e \rangle$ is the average electron density in the system with $N$ number of sites.
	
	The real space Nambu Green's function in our lattice model can be written as \cite{Zhu_2016}, 
	\begin{equation}\label{B7}
		\tag{B7}
		\begin{split}
			\mathcal{G}_{ij}(\tau,\tau^\prime) & = - \langle T_\tau \{ \psi_i(\tau) \otimes \psi^\dagger_i(\tau^\prime) \} \rangle \\
			& = -\theta(\tau - \tau^\prime)  \langle  \psi_i(\tau) \otimes \psi^\dagger_j(\tau^\prime) \rangle  \\
			& \hspace{0.25cm} + \theta(\tau^\prime - \tau) \langle \psi^\dagger_j(\tau^\prime) \otimes \psi_i(\tau) \rangle .\\
		\end{split}
	\end{equation}
	Since the trace is unchanged upon a cyclic variation of operators, one can easily write Green's function as a function of difference in time $\tau - \tau^\prime$, therefore it becomes,
	\begin{equation}\label{B8}
		\tag{B8}
		\mathcal{G}_{ij}(\tau,\tau^\prime) = \mathcal{G}_{ij}(\tau - \tau^\prime) .
	\end{equation}
	Considering the retarded part, i.e. $\tau > \tau^\prime$, of the Green's function, we got  
	\begin{equation}\label{B9}
		\tag{B9}
		\mathcal{G}_{ij}(\tau - \tau^\prime) = \begin{pmatrix}
			-\langle c_{i\uparrow}(\tau) c^\dagger_{j\uparrow}(\tau^\prime) \rangle && -\langle c_{i\uparrow}(\tau) c_{j\downarrow}(\tau^\prime) \rangle \\ \\
			-\langle c^\dagger_{i\downarrow}(\tau) c^\dagger_{j\uparrow}(\tau^\prime) \rangle && -\langle c^\dagger_{i\downarrow}(\tau) c_{j\downarrow}(\tau^\prime) \rangle \\
		\end{pmatrix} .
	\end{equation}
	The Bogoliubov transformation,
	\begin{equation}\label{B10}
		\tag{B10}
		\begin{split}
			& c_{i\uparrow}(\tau) = \sum_{E_m > 0} [u^i_m \gamma_{m\uparrow}(\tau) - v^{i\star}_m \gamma_{m\downarrow}(\tau)] \\
			& c_{i\downarrow}(\tau) = \sum_{E_m > 0} [u^i_m \gamma_{m\downarrow}(\tau) + v^{i\star}_m \gamma_{m\uparrow}(\tau)]
		\end{split}
	\end{equation} 
	BdG eigenvectors $u^i_m$ and $v^i_m$ satisfy $\sum_m (|u^i_m|^2 + |v^i_m|^2) = 1$ for each site $i$ and the sum over eigenvalues $E_m \geq 0 $. The operator $\gamma^\dagger_{n\sigma}(\gamma_{n\sigma})$ create (annihilate) a Bogoliubov quasiparticle of spin $\sigma$ at state $n$ and obeys the anti-commutation relation,
	\begin{equation}\label{B11}
		\tag{B11}
		\begin{split}
			& \hspace{0.5cm} \{ \gamma_{n\sigma} , \gamma^\dagger_{m\sigma^\prime}\} = \delta_{nm} \delta_{\sigma\sigma^\prime} \\
			&\{ \gamma_{n\sigma} , \gamma_{m\sigma^\prime}\} = \{ \gamma^\dagger_{n\sigma} , \gamma^\dagger_{m\sigma^\prime}\} = 0
		\end{split}
	\end{equation} 
	The time dependence of the quasiparticle operators,
	\begin{equation}\label{B12}
		\tag{B12}
		\begin{split}
			& \gamma_{m\sigma}(\tau) = \gamma_{m\sigma} e^{-E_m\tau},\\
			& \gamma^\dagger_{m\sigma}(\tau) = \gamma^\dagger_{m\sigma} e^{E_m\tau}.\\
		\end{split}
	\end{equation} 
	Using Eqs. (\ref{B11}) and (\ref{B12}) in Eq.~(\ref{B9}), one obtains the matrix elements of Nambu Green's function, 
	\begin{equation}\label{B13}
		\tag{B13}
		\begin{split}
			\mathcal{G}_{ij}^{11}(\tau - \tau^\prime) =  - \sum_{E_m > 0} [ & u^i_m u^{j\star}_m f(-E_m) e^{-E_m(\tau-\tau^\prime)} \\
			& + v^{i\star}_m v^j_m f(E_m) e^{E_m(\tau-\tau^\prime)}]
		\end{split}
	\end{equation}
	where $f(E_m)$ is the Fermi distribution function, and we have also used some properties of a statistical average of the quasiparticle operators 
	
	\begin{equation} \label{B14}
		\tag{B14}
		\begin{split}
			\langle \gamma^\dagger_{n\sigma} \gamma_{m\sigma^\prime} \rangle & = \delta_{nm} \delta_{\sigma\sigma^\prime} f(E_n) \\
			\langle \gamma_{n\sigma} \gamma_{m\sigma^\prime} \rangle & = \langle \gamma^\dagger_{n\sigma} \gamma^\dagger_{m\sigma^\prime} \rangle = 0 .
		\end{split}
	\end{equation}
	
	Restricting $\tau - \tau^\prime$ in the range $[0,\beta]$ where the factor $\beta = 1/(k_BT)$, it is straightforward to obtain the component of the Nambu Green’s function in the frequency domain
	\begin{equation}
		\tag{B15} \label{B15}
		\begin{split}
			\mathcal{G}_{ij}^{11}(i\omega_n) & =  \int_{0}^{\beta} d\tau \hspace{0.2em} e^{i\omega_n\tau}\mathcal{G}_{ij}^{11}(\tau) \\
			& = - \sum_{E_m > 0} [ u^i_m u^{j\star}_m f(-E_m) \int_{0}^{\beta} d\tau \hspace{0.2em}  e^{(i\omega_n-E_m)\tau} \\
			& \hspace{0.6cm} + v^{i\star}_m v^j_m f(E_m) \int_{0}^{\beta} d\tau \hspace{0.2em} e^{(i\omega_n+E_m)\tau} ]\\
			& = \sum_{E_m > 0} \biggl[ \dfrac{u^i_m u^{j\star}_m}{i\omega_n-E_m} + \dfrac{v^{i\star}_m v^j_m }{i\omega_n+E_m} \biggr]
		\end{split}
	\end{equation}

	Other matrix elements of the Green’s function can be evaluated in the same way. Finally, in the frequency domain, Nambu Green's function can be written as follows,
	\begin{widetext}
	\begin{equation}\label{B16}
		\tag{B16}
		\mathcal{G}_{ij}(i\omega_n) = \begin{pmatrix}
				\mathlarger{\mathlarger{\sum}}_{E_m > 0} \biggl ( \dfrac{u^i_m u^{j\star}_m}{i\omega_n - E_m} + \dfrac{v^{i\star}_m v^{j}_m}{i\omega_n + E_m} \biggr) &  \mathlarger{\mathlarger{\sum}}_{E_m > 0} \biggl ( \dfrac{u^i_m v^{j\star}_m}{i\omega_n - E_m} - \dfrac{v^{i\star}_m u^{j}_m}{i\omega_n + E_m} \biggr) \\
				\mathlarger{\mathlarger{\sum}}_{E_m > 0} \biggl ( \dfrac{v^i_m u^{j\star}_m}{i\omega_n - E_m} - \dfrac{u^{i\star}_m v^{j}_m}{i\omega_n + E_m} \biggr) & \mathlarger{\mathlarger{\sum}}_{E_m > 0} \biggl ( \dfrac{v^i_m v^{j\star}_m}{i\omega_n - E_m} + \dfrac{u^{i\star}_m u^{j}_m}{i\omega_n + E_m} \biggr) \\
			\end{pmatrix} .
	\end{equation}
	\end{widetext}

	\bibliography{references}
	\end{document}